# μ-PES Studies on TiNCl and Quasi-two-dimensional Superconductor Na-intercalated TiNCl

Noriyuki Kataoka[1]*, Kensei Terashima[2]#, Masashi Tanaka[3], Wataru Hosoda[1], Takumi Taniguchi[1], Takanori Wakita[2], Yuji Muraoka[1,2], and Takayoshi Yokoya[1,2]

[1]*Graduate School of Natural Science and Technology, Okayama University, Okayama 700-8530, Japan*

[2]*Research Institute for Interdisciplinary Science, Okayama University, Okayama 700-8530, Japan*

[3]*Graduate School of Engineering, Kyushu Institute of Technology, Kitakyusyu 804-8550, Japan*

#*Present Address: National Institute for Materials Science, 1-2-1 Sengen, Tsukuba, Ibaraki 305-0047, Japan*

Electron-doped TiNCl is a superconductor, for which exotic mechanisms of the superconductivity have been discussed. However, difficulty in preparing large single-phase samples has prevented the direct observation of its electronic structure and how that changes with electron doping. In this study, micro-photoemission spectroscopy (μ-PES) was used to reveal the electronic structures of TiNCl and Na-intercalated TiNCl (Na-TiNCl). Comparison of Ti 2$p$ core-level spectra shows the enhancement of a spectral feature in Na-TiNCl that suggests an increase in its metallic character. This indicates the introduction of electron carriers to the TiNCl layers. Although the overall valence band electronic structure of parent TiNCl could be reproduced by first-principle calculations, that of Na-TiNCl showed a marked deviation from the rigid band model near the Fermi level ($E_F$). The spectral shape observed near $E_F$ of Na-TiNCl was found to be similar to a result in early transition metal oxides that has been attributed to the effect of strong electron correlation. The present study also demonstrated the usefulness of μ-PES to obtain reliable electronic structure data for samples that are difficult to make at large sizes.

## 1. Introduction

Transition metal nitride halides $MNX$ ($M$ = Ti, Zr, Hf; $X$ = Cl, Br, I) belong to a characteristic class of layered materials that exhibit superconductivity by carrier doping.[1-3] The $MNX$ crystalizes into two polymorphs: α and β forms. The β-$MNX$ has a SmSI-type crystal structure,[4] consisting of a stack of double honeycomb $M$-N layers sandwiched by two $X$ layers, as shown in Figs. 1(a) and (b). Alkali atoms with or without organic molecules are intercalated into the gap between two adjacent $X$ layers by realigning the $M$N$X$ layers. Intercalated compounds exhibit superconductivity with a highest transition temperature ($T_c$) of 25.5 K in the Li and tetrahydrofuran (THF) co-intercalated compound β-HfNCl.[1,2] It has been reported that superconductivity can also be induced by carrier doping with an electronic double-layer transistor (EDLT).[5-7] Electronic structural studies have shown that alkali-metal intercalation adds quasi-two-dimensional electronic states, hybridized states of Hf 5$d$ and N 2$p$ orbitals, near the Fermi level ($E_F$).[8-12] In addition, the absence of the isotope effect, extremely large superconducting gap value $2\Delta/k_BT_c$, and enhancement of $T_c$ for lower carrier concentration strongly suggest an exotic superconducting state in this series of materials.[13-19]

Unlike for the β-form, there are few studies on α-$M$N$X$, which forms an FeOCl-type crystal structure.[20] It possesses stacking of $M$N$X$ layers, including double square $M$-N layers, as shown in Figs. 1(c) and (d). The crystal structure of alkali atom-intercalated TiNCl[3] is also shown in Fig. 1(e). Electron-doped α-HfNBr was considered to be a Mott insulator because electron-doped α-$M$N$X$s with intercalation of Li or Na have been found to have localized electrons down to 2 K.[21] Recently, such alkali-metal and/or organic molecules as pyridine (Py) intercalated TiNCl, were reported as superconductors with $T_c$ = 5-17 K.[3, 22-25] Interestingly, $T_c$ increases linearly with 1/$d$, where $d$ is a basal spacing along the $c$-axis.[23] This is in sharp contrast to the observation in the β-form, where $T_c$ increases linearly with $d$ in electron-doped ZrNCl and HfNCl.[18, 26] The reduced superconducting gap value $2\Delta/k_BT_c$ of $K_x$TiNCl has been reported to be approximately 15, as determined via scanning tunneling microscopy/spectroscopy.[27] The observed value is four times larger than that of the mean field Bardeen–Cooper–Schrieffer value, and it is difficult to explain by the strong coupling theory based on electron–phonon interaction.[28-30] Theoretical studies considering spin and charge fluctuations reported that charge fluctuation is superior to spin fluctuation.[31] More-recent theoretical studies further elaborated the possibility of a charge-fluctuation-mediated model for the superconductivity in electron-doped TiNCl.[32-34]

To understand the pairing mechanism of this possible exotic superconductor, it is important to clarify how the electronic structure evolves with carrier doping. Although there have been several reports of photoemission spectroscopy for the β-form,[10-12, 35] no electronic structure study using photoemission spectroscopy of the parent and doped α-form compounds has been reported. This is because it is difficult to synthesize samples of both the parent and the electron-doped α-form. Moreover, it is not easy to have a uniformly intercalated sample large enough for photoemission measurements. In this study, micro-photoemission spectroscopy (μ-PES) was performed with highly focused synchrotron light to reveal the electronic structure of TiNCl and electron-doped TiNCl via Na intercalation. The overall valence band structure of TiNCl agrees with first-principle calculations while the electron-doped TiNCl has an additional broad state near $E_F$. The observed spectral shape near $E_F$ of Na-intercalated TiNCl (Na-TiNCl) was found to be similar to the result of early transition metal oxides that has been attributed to the effect of strong electron correlation.

## 2. Experimental

The parent compound TiNCl was grown by a method described elsewhere.[3] Na intercalation was carried out using a 0.1-M alkali-metal naphthalene solution in tetrahydrofuran (Na-Naph/THF) in a glove box under an Ar atmosphere. The reactants were mixed so that the resulting compositions had a molar ratio of $Na_{0.25}TiNCl$. Details of this method have been reported in the literature.[23] The lattice parameters of the crystals obtained in this study were determined by the Rietveld refinement to be $a = 3.9373(1)$, $b = 3.2554(1)$, and $c = 7.7997(3)$ Å for pristine TiNCl, and $a = 4.0200(4)$, $b = 3.2709(3)$, and $c = 16.890(2)$ Å for Na-TiNCl. The basal spacing expanded from 7.7997(3) to 8.446(1) Å upon Na intercalation. The composition of the intercalated compound was estimated to be $Na_{0.160(3)}TiNCl_{0.970(3)}$ from the refined site occupation parameters, suggesting a slight deficiency of Cl atoms. These results are in good agreement with those obtained previously.[3,23] As TiNCl and alkali-metal intercalated TiNCl are very sensitive to humid air, they were pelletized in an Ar atmosphere. They were transferred to an ultrahigh-vacuum chamber for photoelectron spectroscopy without being exposed to humid air for a long period.

μ-PES measurements were performed at SPring-8 BL25SU using a DA30 analyzer (total energy resolution ~200 meV). The excitation light diameter was focused down to 10 μm, which is slightly smaller than the typical grain size of the TiNCl. (The size of the microcrystal of TiNCl is approximately 30 μm, as in Fig. 2, and that of Na-TiNCl may be even smaller.) The μ-PES measurement was performed using a photon energy of 1200 eV, and the Fermi edge of Au was used for calibration of the binding energy. To obtain a clean surface, the TiNCl sample was fractured at room temperature under a pressure of $2.6 \times 10^{-8}$ Pa, while the Na-TiNCl sample was fractured at 110 K under the same pressure. As the spectra were position-dependent, especially in Na-TiNCl, a measured position was selected so that the intensity near $E_F$ was at its maximum. From core-level spectra of Na 1$s$, Ti 2$p$, N 1$s$, and Cl 2$p$ at the position, the composition ratio was estimated to be (Na:Ti:N:Cl = 0.2:1:1.1:0.6). The angular dispersion of the valence band states was not observed clearly for both samples. This may be the result of the random in-plane orientation of the samples, because several μ-crystals were illuminated simultaneously, even by the focused soft X-ray.

To examine the electronic structure of the obtained valence band photoemission spectroscopy spectrum, density functional theory calculations for TiNCl were performed using a WIEN2k[36] package with an exchange correlation function proposed by Perdew et al.[37] The crystal structure and $k$-mesh used for the calculation are the same as those in an earlier work.[31] $E_F$ of the calculation was set to the bottom of the conduction band.

## 3. Results and Discussion

This study used core-level data that provided insight into the chemical states of the samples. Figure 3(a) shows Ti $2p$ core-level photoemission spectra of TiNCl (bottom) and Na-TiNCl (top) measured with a 1200-eV photon energy, together with curve-fitting analyses. The spectra were normalized to the area after subtracting the background of an iterative Shirley method. [38,39] In the spectrum of TiNCl (bottom), two peaks were observed at 457.9 and 463.8 eV, which are spin-orbit partners ($2p_{3/2}$ and $2p_{1/2}$) of Ti $2p$, respectively, with a tail toward the lower-binding-energy side. The binding energy of the peak matches well the reported value of Ti 4+,[40] indicating that the dominant Ti valence state of TiNCl is 4+ from an ionic picture. In contrast to the Ti $2p_{3/2}$ of TiNCl, that of Na-TiNCl is relatively wide, having a new structure at approximately 455 eV. This energy agrees well with the Ti $2p_{3/2}$ binding energy of metallic TiN,[41] suggesting an increase of metallic character as a result of charge transfer from the Na site to the TiNCl layer in Na-TiNCl. The spectral change is different from the observation of operando X-ray photoemission spectroscopy core-level study of β-ZrNBr, where carrier injection using EDLT does not lead to drastic change in the Zr $3d$ spectral shape.[42]

To gain further insight into the chemical states, fitting analyses were performed for the Ti $2p$ spectra of TiNCl and Na-TiNCl. The Ti $2p$ region was curve-fitted to pairs of spin-orbit doublet peaks with a fixed area ratio of 2:1 between each Ti $2p_{3/2}$ and Ti $2p_{1/2}$ component. The energy separation of the spin-orbit partners used ranged from 5.7 to 5.9 eV, depending on Ti valence. A wider line was used for Ti $2p_{1/2}$ compared with that for Ti $2p_{3/2}$; this was caused by Coster–Kronig broadening.[43] The spectra were fitted using the minimum number of components and changing their energy position, energy width, and intensity. At least three components were necessary to reproduce the TiNCl spectrum, as shown in Fig. 3(a). While the higher-binding-energy component lies within the energy region of Ti oxides, the lower-binding-energy component is within the energy region of compounds with intermediate

Ti valences between 3+ and 4+.[44)] For Na-TiNCl, the two components G and H, which correspond to components B and C, respectively, were used to fit the TiNCl but shifted toward a higher binding energy. Then the two components D and F, which are attributed to the main and satellite components, respectively, were added by referring to the spectral components of TiN.[44)] However, it was difficult to reproduce the spectral shape. Therefore, another component, E, was added to fit the spectrum, which suggests the existence of intermediate valence states between Ti 4+ and 3+.

Figure 3(b) shows the Cl $2p$ core-level spectra of the two compounds normalized to the spectral intensity of the Ti $2p$ core level. It was found that the intensity of Cl $2p$ in Na-TiNCl was somewhat smaller than that in TiNCl, indicating that Cl atoms are less contained in Na-TiNCl. Such a removal of Cl atoms, which also provides electron carriers to the system, was reported in β-*M*NCl and denoted "Cl deintercalation."[45)] Therefore, it is possible that there was an additional doping caused by Cl deintercalation in the sample. From the spectral analyses, it was found that the number of components to reproduce the spectrum was larger in Na-TiNCl than in TiNCl, as shown in Fig. 3 (b), indicating increasing chemical components for Na-TiNCl. The higher-binding-energy components that may have been formed by reaction with the solvent used for the intercalation of Na were located within the energy range of Cl $2p$ in organics.[46, 47)]

The blue line in Fig. 4(a) shows the valence band photoemission spectrum of TiNCl compared to the results of band structure calculations. The experimental spectrum has a peak of approximately 6 eV and a shoulder structure of approximately 4 eV. The intensity decreases for a lower binding energy and with negligible spectral intensity in the energy range from 2 eV to the Fermi level ($E_F$). By comparison with band structure calculations, it was found that the experimental valence band can be explained with calculated bands that are shifted by 1.3 eV to the high binding energy. Thus, it can be concluded that the 6-eV peaks are hybridized states of Ti $3d$, N $2p$, and Cl $3p$ orbitals and the 4-eV shoulder structure is dominated by the Cl $3p$ component.

In the valence band photoemission spectrum of Na-TiNCl shown in Fig. 4 (top), a broad peak was observed at 6.5 eV and a smaller structure at approximately 0.5 eV. From the similarity of the spectral shape, the former is attributed to the hybridized states of the Ti $3d$, N $2p$, and Cl $3p$ states. As for the valence bandwidth, Na-TiNCl has a wider bandwidth than TiNCl, as evidenced by the observation that the spectral weight of Na-TiNCl ranges from around 2 eV to approximately 9.5 eV, while that of TiNCl covers from 2.2 to 8.5 eV. This broad structure of Na-TiNCl has a higher binding energy by 0.5 eV than the corresponding band of TiNCl, in line with the chemical potential shift expected from electron doping. However, the new structure at $E_F$ extends to 1.5 eV. This observation contradicts a previous study that reported that the shapes of the valence and conduction bands of Na-TiNCl are similar to those of TiNCl.[3] It has been suggested that it is difficult to explain the observed valence band structure of Na-TiNCl, especially the spectral structure near $E_F$, only by the rigid band shift of that of TiNCl.

To gain further insight into the electronic structure of Na-TiNCl, the near $E_F$ region of the spectrum of Na-TiNCl was expanded, and it is shown with a calculated photoemission spectrum in Fig. 5. The calculated spectrum was made from the calculated band structure, considering the photoionization cross-sections,[48] the Fermi–Dirac function, the experimental energy resolution, and the composition ratio estimated from the core-level spectra. For the doped-electron concentration, two cases were considered (case A: Na intercalation and case B: Na intercalation + Cl deintercalation). In the experimental spectrum, a Fermi edge was observed, consistent with the metallic nature of the sample. A structure centered around 0.5 eV was also observed that extends to ~1.5 eV. The calculated spectrum of case A reproduces a small peak at $E_F$, although the area of the peak is much smaller than that of the experiment. This suggests that the electron concentration calculated by taking only Na intercalation into account is too small. In case B, a sharp peak was found at $E_F$, and it was found that the total spectral area of the peak was the same order as that of the experiment. This suggests that the electron concentration of the experiment is more consistent with that expected from Na intercalation + Cl deintercalation. The width of the peak near $E_F$ observed in the experiment was, however, much broader than that of the calculated one.

There may be several reasons for the broader experimental spectral shape compared with the band calculations. One of the reasons may stem from the distribution of sample composition, probably the distribution of doping concentration. This is in line with the broader valence bandwidth and multiple core-level components. Then, the use of μ-PES helped to reduce this effect. Indeed, a strongly position-dependent spectral shape was observed in Na-TiNCl, in contrast to an almost position-independent valence band shape in the parent TiNCl. Another possible reason is that there could be two sources of electron doping for the present sample, Na intercalation and Cl deintercalation, as the core-level analysis shows. While electron carriers by Na intercalation may occupy the conduction band of TiNCl with a rigid band manner, electrons doped by Cl deintercalation may be operating locally. Actually, in electron-doped β-$M$NCl, the observed superconducting properties of the Na-intercalated and Cl deintercalated samples were reported to be different.[9, 27-29,49, 50] This observation supports the idea that the effects of electron doping may differ depending on the source of the electron doping, although the observed difference may also be explained by the difference in the basal spacing $d$.[28, 29] Finally, it is interesting to recall that the doped α-HfNBr system has been discussed as a possible Mott insulator, because the band calculation failed to predict its insulating behavior.[21] The broader experimental spectral shape of the states at $E_F$ compared with that of the calculation has similarly been observed in early transition metal oxides and attributed to the effect of strong electron correlation.[51] The similarity suggests that Na-TiNCl may be a correlated metal, having an incoherent part of spectral function in the higher binding energy side of the coherent part in the vicinity of $E_F$. If this is the case, the present observation can help to understand the reported novel superconducting properties.

## 4. Conclusion

In summary, μ-PES was performed to reveal the electronic structures of TiNCl and Na-TiNCl. A change in the spectral shape of Ti $2p$ was observed; this suggests an increasing metallic character in Na-TiNCl. The shape of the valence band of TiNCl corresponds roughly to that of band structure calculations. The valence band spectrum of Na-TiNCl exhibited a peak at 6.5 eV, accompanied by an emergence of the metallic state near $E_F$ that extended to 1.5 eV; this is different from that expected from band calculations. Similar observations have been made in other correlated early transition metal oxides. The present study also demonstrated the usefulness of μ-PES for obtaining reliable electronic structure data for samples difficult to make in a large size.

**Acknowledgment**

We thank Professor Yoshihiro Kubozono and Tomoya Taguchi for support in preparing the samples. We would also like to thank Professor Shoji Yamanaka of Hiroshima University for his helpful suggestion on the synthesis of samples. Experiments at SPring-8 were performed under proposal numbers 2018B1283 and 2018B1369. This work was partially supported by a Grant-in-Aid for Scientific Research on Innovative Areas "3-D Active-Site Science" (No. H1705220) and by the Fund for the Promotion of Joint International Research (B) (No. 18KK0076) from the Ministry of Education, Culture, Sports, Science, and Technology of Japan (MEXT). The upgrade of BL25SU for reducing the spot size of the light was supported by the Photon and Quantum Basic Research Coordinated Development

Program from MEXT.

*E-mail: po5n3yf0@s.okayama-u.ac.jp

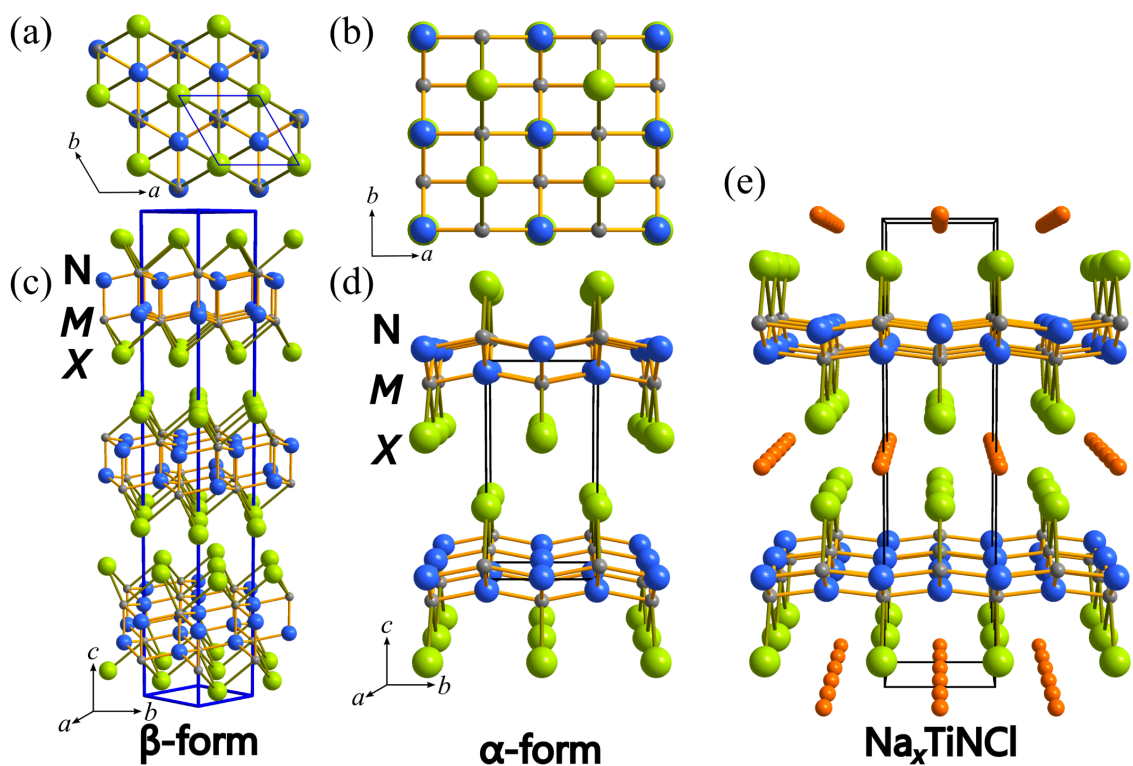

Fig. 1. (Color online) Crystal structure of MNX: (a) and (b) are top and side views of β-*M*N*X*, respectively, (c) and (d) are the top and side views of α-*M*N*X*, respectively, and (e) is the side view of Na-intercalated TiNCl. Upon intercalation, TiNCl shows polytype shift, and the lattice type changes from the space group *Pmmn* of the pristine TiNCl to *Bmmb*.

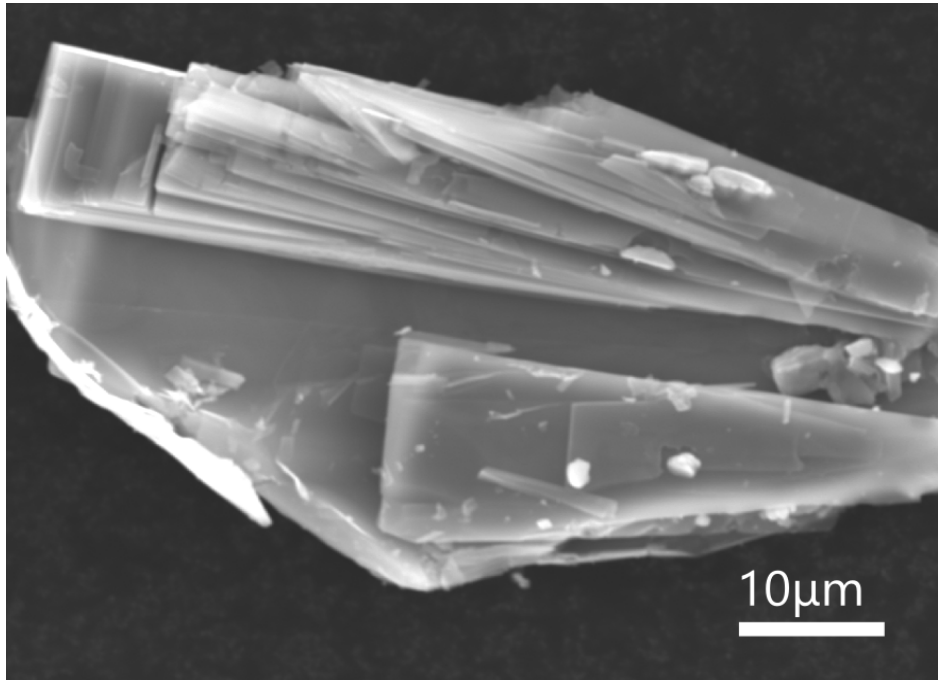

Fig. 2. Scanning electron microscope (SEM) image of TiNCl single crystal.

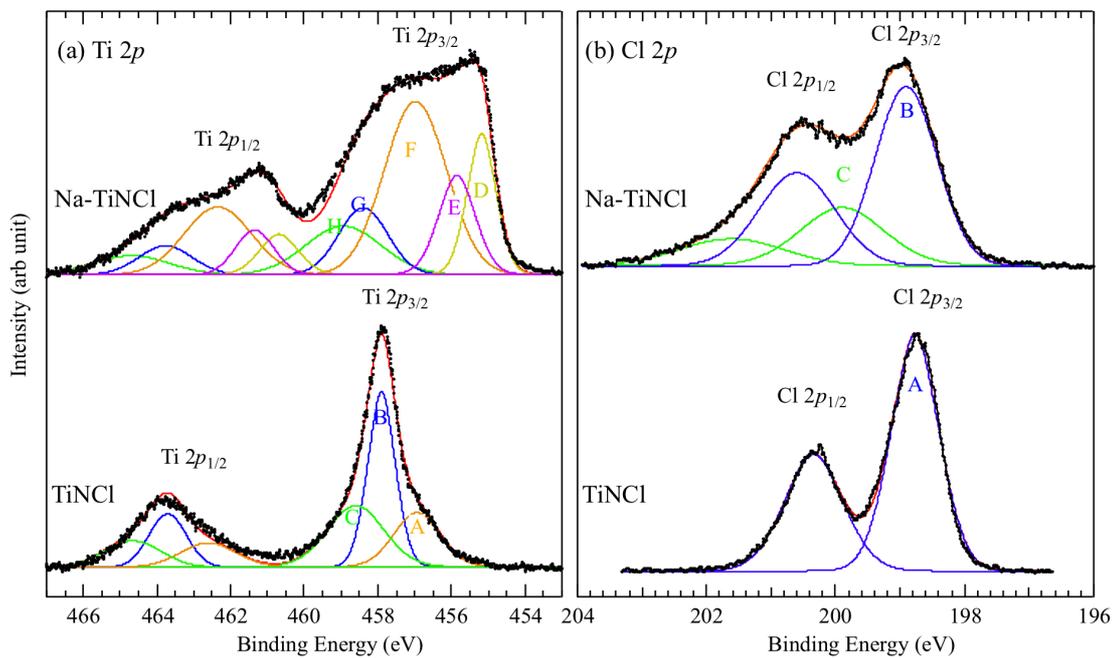

Fig. 3. (Color online) (a) Ti 2p core level spectra and (b) Cl 2p core level spectra of TiNCl (bottom) and Na-TiNCl (top) measured with photon energy of 1200 eV. The results of the fitting analyses are also shown with curves.

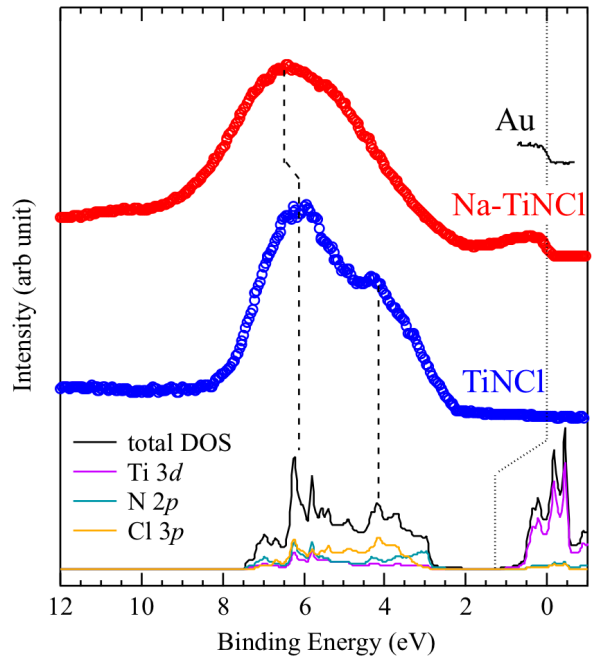

Fig. 4. (Color online) Valence band photoemission spectra of TiNCl (blue open circles) and Na-TiNCl (red open circles) measured with photon energy of 1200 eV, together with calculated total and partial density of states of TiNCl. Whereas $E_F$ of the experimental spectra is determined with the Fermi edge of a gold film electrically contacting the samples, $E_F$ of the calculation is set to the bottom of the conduction band and is shifted to 1.3 eV toward the higher binding energy side, as indicated by the dotted line.

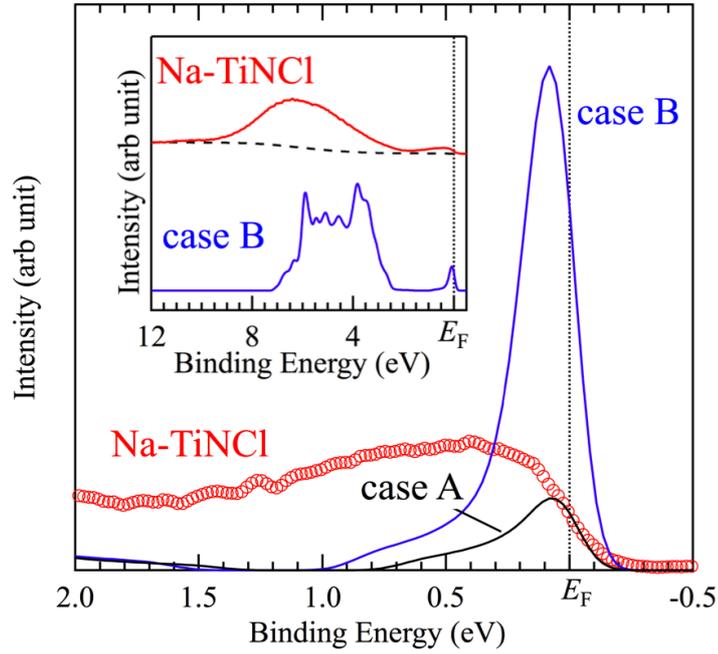

Fig. 5. (Color online) Near-$E_F$ photoemission spectrum of Na-TiNCl (red open circles), simulated PES spectrum of case A (Na-intercalated TiNCl, black solid line) and case B (Na-intercalated + Cl deintercalated TiNCl, blue solid line), with the valence band spectra (inset). Normalization of the spectra was done with the total intensity of the valence band spectra, after subtracting a background of iterative Shirley (a broken line in the inset).